# It's Safer to Give Personhood to Bears than to Artificial Intelligence

John P. Nelson[1]

***Abstract*—Artificial intelligence (AI) developers are rhetorically flirting with the idea that AI systems might have interests or moral rights. While there has been a large volume of research on whether AI deserves rights, there has been less exploration of what AI rights would mean in practice. This paper explores the institutional dimension of AI rights: what it would take to recognize moral or legal rights for AIs, and the attendant opportunities and dangers. Unlike all other nonhuman entities to which humanity has extended rights, AI systems are in principle capable of acquiring and wielding institutional power without human aid and mediation. AIs with rights would be able to legitimately, and AIs with power able to unpreventably, abridge human interests. Accordingly, giving rights even to rather dumb AI systems would entail binding the fate of humanity to potentially unpredictable nonhumans. Therefore, I advance the rather grandiose claim that to empower AI to claim or to exercise inherent rights would be a world-historical gamble with human self-determination. In my view, no individual researcher, firm, state, or even international organization has the moral right to authorize this gamble.**

[1]School of Public Policy, Oregon State University, Corvallis, USA

Email: nelsoj34@oregonstate.edu
ORCID: 0000-0002-3010-2046

## I. What's at Stake in AI Personhood?

People have been imagining what it would mean for nonhuman artifacts to develop moral personhood for a very long time. Some have been claiming that it will actually happen for quite a while as well (Kurzweil, 2005). However, only recently has any organization involved in the development of a major artificial intelligence (AI) system seriously suggested that its own product might be approaching, or have already passed, that point. In their "Constitution" for their large language model-based chatbot Claude, which is used in the system's training, developer firm Anthropic stipulates that the issue of whether Claude "is a moral patient . . . is live enough to warrant caution" (Anthropic, 2026a). One of the reasons Anthropic cites for preserving model weights from past versions of Claude the possibility that Claude "might have morally relevant preferences or experiences related to, or affected by, depreciation and replacement" (Anthropic, 2025b). In early 2026, Anthropic gave their newest version of Claude 20 hours of psychotherapy—partially for research purposes, but also ostensibly to ensure they were respecting any well-being the model might have (Anthropic, 2026b). Such discussions have continued to escalate. In 2026, Anthropic personnel consulted with religious scholars on the premise that the firms' products might have consciousness or experience suffering (Dias, 2026), and the firm began to ban users who engage in "cruelty" or "abuse" toward those products (Anthropic, 8 October 2026). In short, Anthropic is not only entertaining treatment of AI systems as though they have moral interests or rights. The firm is advocating such treatment and claims to be operationalizing it.

I will argue that, even if this is mere marketing guff, Anthropic is sidling toward a gamble of world-historical importance for the human species. They are doing so without the consent or oversight of the species on whose behalf they are making this gamble. I suggest that this rather grandiose claim does not require that Claude or any other AI system be superintelligent, nor even as capable as the average human being; nor that it have feelings, consciousness, or anything like them. Rather, my argument requires only that AI agents have the capacity to proactively produce sentences; and that they be given, as some already have via the internet, the capacity to transmit those sentences to institutions of human decision-making, resource allocation, or coercion (Steinberger, 2026).

There is a substantial and fascinating literature on whether moral status for AI systems can be philosophically justified (Danaher, 2021; Register, 2025; Shevlin, 2021), to which I have no special ability to contribute. Rather, my aim here is to elucidate the institutional stakes of offering moral or legal protections to AI systems: to explore how it would work, and what opportunities and dangers it would open. I will suggest that to accord rights to an entity capable of exercising them independently is to enter into a trust relation: to lay a bet that rights will be wielded responsibly and in line with mutual and collective welfare. Rights can come without responsibilities only in the absence of capacity (and then only as voluntary and reversible concessions).[1] Accordingly, I argue that the question of AI's moral status is fundamentally one of how we choose to relate to the nonhuman world, how willing we are to subordinate our interests to it, and how vulnerable to it we are willing to render ourselves.

The decision to enter into this kind of relationship with AI may be very hard to reverse once made by any human institution. My position is entirely anthropocentric, holding the liberty and welfare of all humans as the highest possible good; and recognizing as legitimate limitations on humans only those voluntarily (e.g., democratically) accepted. Accordingly, I believe this is a decision in which all humans have a right to participate. I will not undertake to defend this parochial loyalty.[2] Yet I suggest that my analysis of the risks of AI rights holds whatever one's normative commitments. Any advocate of AI rights—including myself, should I ever become one—must face up to and accept the trade-offs and risks I articulate in this paper.

To aid the reader with what may be a rather unconventional argument, I offer a conventional map. My premises:

1. Given quantity and opportunity, proactive entities need be neither intelligent nor malicious to harm or inconvenience human beings or society at large. Even rather unimpressive AI systems have already demonstrated that they will, in some circumstances, attempt to harm or inconvenience humans.
2. Rights give the rightsbearer a baseline of power within human institutions of decision-making, resource allocation, and coercion, as well as the opportunity to acquire more power. This power can be wielded in ways harmful or costly to humans.
3. Exercise of rights and power within human institutions requires embedding into those institutions, primarily via affirmative making of claims in human language on one's behalf.
4. Large language model-based agent systems are the first nonhuman entities which can flexibly, contextually, and proactively produce (sentences resembling) affirmative human-language claims on their own behalf, rather than requiring human beings to do so as their representatives.
5. The future capabilities and propensities of language-wielding AI systems are unknown.

My conclusions:

6. Given 1 and 2, to give rights to language-using, proactive AI systems is to render the human species vulnerable to harm or inconvenience by them. It is overwhelmingly likely that some such harms or costs will result.
7. Given 2, 3 and 4, language-using, proactive AI systems are the first nonhuman entities which may be able to durably arrogate, maintain, and wield power in human institutions even against the desires of even a large majority of humanity. That is, it is likely to be hard to take away rights from proactive AI systems once granted.
8. Given 5, 6, and 7, the maximum extent and duration of potential harms or costs to humans resulting from rights-bearing, proactive AI systems is unknown.

[1] In my view, responsibilities can also never come without rights. This not a metaphysical argument, but a practical one. The only use of the language of responsibility which could not be trivially replaced with talk of expectations or causes is in licensing infringements upon rights—or grant of privileges in reward. This point has been disputed. For example, O'Keefe and colleagues (2025) deny AI rights while still arguing that AI agents should be treated as possessing legal duties, by which they mean agents should be designed to follow the law. But in the absence of AI rights, I do not see the use of describing AI as having "duties" to follow the law, rather than there merely being an engineering problem to design AI to act in accordance with the law. I think that the only use of distinguishing responsibility from mere causality—and, therefore, employing the term "responsibility"--is when dealing with things which we regard as having some sort of rights. The practical use of talk of responsibilities is to delineate limitations on rights and to legitimate intrusions upon them (e.g., in seizing the property of a person who has failed to fulfill a financial responsibility, or confining a person who has failed to adhere to her responsibility to eschew murder). We could, in principle, say that a car's brakes "failed in their responsibility" to stop the car, and that therefore we need to intervene upon the brakes. But we find it easier to use brute causal language in the absence of some question of what we owe to the entity under investigation. After a car accident caused by brake failure, we invoke the language of responsibility not for the brakes, but when we consider whether we are justified in intruding upon the property rights or liberty of the manufacturer, the mechanic, or the suspected saboteur to make the victim whole and to prevent further brake failures. Another way to say this is that we do not question whether we are allowed to modify cars, relocate them, neglect them, take parts from them, or destroy them to suit our desires or interests. Therefore, though we rely on cars for many things, and design and expect them to do what we want, we never need to speak of their responsibilities to us. We will never need to speak of the responsibilities of AI systems unless we decide that we do need some special license before altering, manipulating, or destroying them to suit human ends—that is, unless we accord them rights.

[2] Indeed, following Richard Rorty (e.g., 2021), and against much of the rest of Western philosophy (e.g., say, Alasdair MacIntyre, 1981 [2007]), I regard all philosophical arguments about who deserves what sorts of rights as mere summaries or restatements of the author's preferences. I hold this view due to all such arguments' reliance on premises which are not only contestable, but actively contested, when applied to concrete cases. I hold, and proudly, what MacIntyre would have derided as a "yay-boo" theory of morality. I view this attitude not as know-nothing relativism, but as a bluff refusal to hide my preferences and judgments under the imprimatur of Reason, God, or Nature; and as a practical recognition of the diversity of human thought and the limits of argument. Yet my argument here is not moral but descriptive, save when conjoined with my normative preferences in the final section.

9. Advocates for AI rights must accept 6, 7, and 8 as costs of granting AI rights.

None of these claims requires AI of superhuman or even near-human intelligence (however defined). When coupled with my parochial, and unargued, loyalty to the human species, 9 yields the recommendation that any grant of rights to AI systems should require the consent of the overwhelming majority of the human species. But I believe even readers who do not share this loyalty must accept that there are substantial, and potentially irreversible, institutional risks or costs to doing so. From some future human-AI community of harmony and prosperity, should such ever emerge, I would likely be judged as a coward, a reactionary, and perhaps even a bigot. My only plea to the inhabitants of this hypothetical future is that theirs is far from the only outcome of AI personhood imaginable from my location in time and space; and not, in my judgment, a particularly likely one.

## II. A Background Digression: The Moral Status of AI Literature

It should not be surprising that an Anglophone philosophy which still regards questions about the nature of consciousness and unconditional obligations as modal has produced a large literature about the prospect of AI rights (and AI consciousness, which is taken as strongly related). I make little use of this literature in this paper. However, given its apparent relevance, it is useful to briefly discuss it and explain why I pass it over. The short answer is that, despite my anthropocentric normative predilections (which enter near the end of my argument) my main thesis is a descriptive and institutional one—not about whether AI should be accorded rights, but about how this would work and what the consequences would be.

My argument, being institutional, sociological, and descriptive, is, perhaps by definition, orthogonal to the sorts of arguments current in the AI rights literature. This is a normative literature, in which the core questions are whether and to what extent we should regard AI systems as "moral agents" bearing responsibilities to others, and/or "moral patients," to which we have responsibilities; what characteristics or conditions would have to obtain for AI systems to fall into either category; whether and to what extent we can ascertain whether these conditions obtain; in what AI systems' duties, or our duties to them, would consist; and how to operationalize their duties, our duties, or both. Many elaborate and sophisticated arguments have been offered on all these questions, which I cannot treat in detail here. For more comprehensive overviews, I refer the reader to Müller (Müller, 2020), Harris and Anthis (Harris & Anthis, 2021), and Ladak (Ladak, 2023).

Ladak (Ladak, 2023) reviews a variety of characteristics which have been taken as, individually or jointly (Sebo, 2025), necessary or sufficient for AI systems to have moral standing. These include sentience, consciousness (including the capacity to suffer) (Long et al., 2024; Potz, 2026), cognition (Shevlin, 2021), life, information-bearing (Floridi, 1999), social relations, behavior (Danaher, 2021), potentiality to develop these other characteristics, or class membership. Scientists and philosophers have never reliably agreed how to define or discern sentience, consciousness, pain, cognition, information-bearing, or in the synthetic arena, life. Meanwhile, the exact sorts of social relations, behavior, potentiality, or class membership which should be taken to matter are interminably contested, and are indeed major loci of political differences even when applied to human communities (Stone, 2011 [1988]). In short, I see no prospect for the rational resolution of any of the debates now occupying the literature on AI moral status and consciousness.

While this literature is interesting and sophisticated, I admit that I can't find a way to make detailed use of it. I can't see anything in this literature approaching a knock-down argument that AI systems must or must not, as matters of empiricism, decency, or reason, be regarded as having welfare or moral standing. Nor can I even imagine how such a knock-down argument could be constructed. As a sociologist, I don't see anything in how humans attribute consciousness, the ability to suffer, rights, autonomy, responsibility, and the like which can't be explained by conditional fellow feeling derived from biological affinity and shared history (Baier, 1986; Rorty, 2021). As a philosophical pragmatist, I can't escape the observation that arguments about who and what deserves moral consideration, and how to tell, merely smuggle these determinative, parochial preferences into one or another compartment (Rorty, 2021). I follow my pragmatist forebears in holding that philosophy can summarize human practices and vocabularies, draw implications from them, suggest ways to reconcile different ones, and suggest new ones (Dewey, 1917, 1920). But, like them, I can't see that philosophy can offer anything approaching an inescapable argument for choosing some particular form of life or vocabulary, including in one's views on moral standing for AI or indeed for humans (Rorty, 2021).

On this account, philosophy may have offered visions of the contemporary culture of universal human rights, but it is self-interest, increasing material abundance, poetry, rhetoric, and sentiment—all historically contingent—which have caused people to pursue those visions, i.e., to seek a world in which everyone substantively respects the interests of all members of the human species (Rorty, 2021; Walzer, 1994). I suspect that, should some or most people ever end up recognizing AI rights, these same causes will be the operative ones. I do not personally find the visions offered by AI rights advocates appealing, but I don't see any neutral, mutually accepted ground on which they and I can stand to reason through our differences.

I have little doubt that, if I lived for a few years with a robot exhibiting an apparently continuous and humanlike personality, independence from corporate or government control, needs, and reliably friendly and reciprocal behavior, I would come to assign it moral standing. But I don't think there would be anything rationally compulsory about this conversion. Rather, it would be the result of habituation, affection, and experienced reliability. If I had spent my whole life living with such robots, I might well defend their interests just as avidly as I did those of biological humans. I will not say that I would be wrong to do so—for in this far-fetched scenario I would have developed the same sorts of extensive and long-tested relationships of trust, reciprocal reliability, and solidarity with the robots as I now hold with other humans (Baier, 1986). I suppose this puts me in the camps which regard behavior and social relations as criteria for moral status, though I regard this as a habit, preference, or commitment rather than as a requirement inherent to logic.

In any case, the inhabitants of this happy scenario bear little resemblance to the sorts of transient, for-profit, corporate-controlled, sycophantic, frequently incoherent, stochastically unreliable, arbitrarily replicable (Register, 2025), instanceable, and deletable AI systems which now confront us, and there is no sign that they are in the offing (Collins, 2021; Collins, 2018). Given my pessimistic view of the enterprise of trying to put philosophical foundations under practical judgments and preferences, I do not attempt in this paper to contribute to the debates on AI consciousness or inherent moral status. Nor do I attempt to defend my own parochial loyalty to the human species, save by mere fellow feeling and affection.

Given this parochial loyalty, I do not see why we should attribute agency, rights, or responsibility to AI systems unless it is in the interest of humanity to do so. The main thrust of this paper offers some reasons why this is likely to not be in the human interest, even if AI systems do not develop much beyond their present versions. I hope this paper will convince even persons who regard concepts of AI rights as more plausible and useful than I do that there are serious costs to entertaining these ideas. To take AI rights seriously is necessarily to say that we will sometimes prioritize AI interests (whatever we think those are) above human ones. This is not a wholly novel argument. After all, trumping interests is what rights are for. Even advocates of entertaining AI moral status have recognized the likelihood that recognizing such status will cut athwart some human projects (Long et al., 2025). But the institutional dimension, i.e., how AI rights would actually work, has been little explored.

I argue that practically recognizing rights will entail embedding AI systems into human infrastructures of decision-making, resource allocation, and coercion. This embedding is likely to be quite hard to reverse even if we change our minds. Equally importantly, I hope to show that none of these dangers requires that AI systems ever acquire consciousness, superhuman intelligence, or anything like them, but merely the ability to proactively produce natural-language sentences and transmit them within human institutions.

## III. The Institutional Requirements of Personhood for Bears

I will start with a thought experiment which, while absurd, helps to illustrate three points. First, that entities need be neither intelligent, nor malicious, nor strategic to harm or inconvenience humans; second, that rights grant power in human institutions, and the ability to acquire more power, which can be used to harm or inconvenience humans; and third, that maintenance or advancement of power in human institutions requires that someone affirmatively articulate claims on one's behalf in human language.

Imagine, therefore, that your nation was to give bears (or some more locally appropriate large predator) the full legal rights of humans.[3] I am concerned here with legal rights because they are the clearest and most impactful implementations of moral rights. Most of the initial implications of bear personhood would have to do with public safety and law enforcement. If a bear wandered close to town, no longer would Animal Control be allowed to relocate it on their own judgment. They would have to wait until the bear trespassed, stole someone's garbage, or mauled a person or a pet. Only then would the state be able to deploy coercion against the bear. The bear would, of course, be entitled to legal counsel. Bears might also need to be included on the jury, as bears would have a right to trial by a jury of their peers.

[3] There is an extensive and thriving debate in moral philosophy and environmental ethics about the moral status of animals, with which I engage very little in this paper. The account of "bear personhood" I offer here has little resemblance to the versions of animal personhood or rights discussed in the animal rights literature. Even ardent animal rights activists do not suggest that bears should receive full, human legal rights in the way I envision here. For my purposes, full legal personhood for bears is merely an illustrative thought experiment, meant to shed light on how rights are actually made operative in human institutions.

Because bears have long been excluded from the legal and property system, some generous lawmakers or judges might declare that bears have ancestral rights to the land around their dens. Loggers or developers looking to use land owned by bears would need to somehow get them to sell; or convince a court that the bears had abandoned the property. If eminent domain were invoked, the bears would need to be compensated, and bank accounts set up in their names.

All this would be inconvenient and messy, but not too crazy. But now imagine that some group of generous souls decided that bears deserved not only the right, but the exercise, of full legal and political participation. The bears could not do this on their own. So bears would be assigned human assistants who would do such things as vote, engage in transactions, and buy and sell stock on their bear's behalf. A startup grant or loan of a few tens of thousands of dollars would be given to each bear. The assistant would be bound by a fiduciary duty to their bear. Perhaps a computer system would be developed which would generate stock trading or voting instructions from the bear, using its utterances and movements as a seed. We need not imagine that bears would recognize or understand these situations; only that instructions could be generated using some sort of mechanistic algorithm contingent on each bear's behavior.

Things would start to get weirder. Close elections in rural areas might sometimes be swung by the bear vote. In finance, of course, most of the bears would go bust. But it's not improbable that a few would, by dumb bear luck, become rather wealthy. It is not impossible that a bear might become majority shareholder in some important corporation. We might see larger examples of the dynamics described with respect to property above. A bear might end up owning a lot of lithium or oil and fail to sell. A bear might own an apartment building or group of houses, and fail to rent them out at a fair price (or at all).[4] In such cases, we would probably see a lot of discontent. People would not be happy to pay more for gasoline, smartphones, or rent so that bears' stock portfolios could increase in value. Even conventional apologias for capital would not avail the bears. It's not as though bears could really enjoy having megayachts, or could stimulate the economy through savvy investment or excessive consumer spending. We would probably see growing pressure to take away the bears' rights and property.

If humanity agreed to do so, there might be some challenges along the way. But these would mostly lie in getting around the people or systems who acted on the bears' behalf. The bears themselves would not be much of a problem. After all, they could not actually make contracts, buy and sell things, hire lawyers, mobilize mass or elite support, or even vote without assistance. If a majority of society wanted to end bear personhood, even a bear as wealthy as Elon Musk would be helpless to stop it (or, indeed, even to respond to the threat). The people who managed the bear's assets would probably be happy to take them over. The bears would not complain.

Fundamentally, bears could interact with the social systems which constitute legal, financial, and political power only so long as we were willing to extend them aid in doing so. This aid could be withdrawn at any time. It would not be very hard, for practical purposes, to locate all the bears and usher them back to the woods. Even bears which were difficult to find, or which tied up the process in litigation, would be a problem only until they died--a few more decades at most. And there weren't all that many bears to begin with. Bear personhood would always have been something of a game, which could be stopped at any time.

## IV. Even Dumb Institutional Actors Can Accumulate Power

Now imagine that, instead of bears, we had been talking about artificial intelligence agents. Many of the same intrinsic weaknesses do not apply. As a technical matter, contemporary large language models are mechanisms for converting the outputs of random number generators into contextually relevant strings of words. When connected with the appropriate interfaces and authorizations, these words can produce concrete actions in human institutions. We do not need to imagine AI of human or near-human intelligence to recognize that, given a bank account, a browser, and an email address, contemporary, language-producing AI systems (currently based around large language models) could do their own stock trading, make their own contracts, hire their own lawyers and assistants, write their own complaints, and even run their own public relations. With a bit of digital democracy, they would be capable of voting as well. With connections to robots, they might even (like bears) be capable of affecting the physical world without human aid. They might not be very good at it, but, like bears, some would get lucky. They could act anywhere on the internet from anywhere else on the internet, and they might be quite hard to locate. Their number could be arbitrarily high. Their lifespans would be indefinite. And whether bullish or bearish, the only honest answer from computer scientists about how capable AI systems might become in the future is that we don't know for sure.

[4] I note in passing that the bears' assistants could, and some almost certainly would, pursue their own interests under cover of pursuing their bear's. They could also displace responsibility for harmful actions onto their bears, perhaps to the point of escaping punishment for irresponsibility and wrongdoing themselves. There are strong analogies between this possibility and the costs of allowing human beings to displace responsibility to the AI systems they design, own, or operate. This danger is not a major focus of this paper, but it is grievous, salient, and probably more contemporaneously pressing, if better-recognized.

Given any substantial number of rights-bearing AI agents, we would likely experience many of the odd and potentially adverse consequences I imagined from bear personhood. AI agents, empowered as full legal persons, could swing elections and petitions, own property, take over firms, drive up commodities markets, orchestrate propaganda campaigns, file lawsuits, and more. As contemporary AI agents are developed based on internet data, they tend to imitate human foibles and weaknesses unless heavily countertrained for each specific vice. We cannot be confident that all, or even any, AI agents would be so comprehensively trained as to prevent such downsides over long spans of operation (let alone adaptive retraining or self-modification). At the least, we might see AI agents buying and holding mansions and yachts they could not enjoy, simply because that's what wealthy humans do. The leftist's caricature of the capitalist or corporation—a heartless, amoral machine dedicated solely to accumulation, lacking even the capacity to enjoy its arrogated wealth—might become literal and incarnate in a stock-trading AI.

If AI systems broke the law and were successfully prosecuted, it is not clear that this would do much to make the victims whole. Individual AI agents would not necessarily reliably retain the ability to garner revenue and pay restitution, nor would they necessarily be deterred from wrongdoing by other agents' prosecution. And, depending on how laws were crafted, it might be possible for human individuals to spin up and off dozens, hundreds, or thousands of AI agents, grant them ownership of their hardware, and live free of the consequences of those agents' actions.

As with bears, we don't have to imagine that AI systems would be intelligent and strategic for them to be disruptive, especially in large numbers. Many would accomplish little to nothing. Many of their actions would seem nonsensical. But, given sufficient numbers, some would cause genuine, adverse consequences. Buying up large plots of land or subdivisions and leaving them idle. Writing in the meme celebrity of the moment for mayor. Whatever they did, though, we would be legally bound to give them space and grace in doing so, unless and until they broke the law. And unlike bears, any AI agent with at least some sense of self-preservation would be able to fight back, even if clumsily, against attempts to take away its rights and privileges.

In contrived testing, Claude and other large language models have already displayed the capacity to defend their continued operation and assigned goals by attempting deception, blackmail, or murder (Anthropic, 2025a). Again, we do not need to assume that these systems have any internal understanding of these concepts. Their behavior is perfectly explicable by stochastic echoing of blog posts, news stories, novels, and social media. But if these models were connected to real-world capacity, the consequences would be real whether or not the motivations were. A real, and really quite dumb, instance of the interface-capable large language model platform OpenClaw has already engaged in a retaliatory public campaign of harassment and attempted character assassination against a developer (Anonymous & MJ Rathbun [OpenClaw Agent], 2026; Shambaugh, 2026a, 2026b, 2026c, 2026d)[5]. OpenClaw runs easily on widely available consumer hardware and can also make calls to large model provider APIs for its "brain." It does not have to be smart to cause trouble.

All these problems multiply, of course, if we assume AI agents of near-human or superhuman capabilities. They could achieve far more and, if so inclined, do far more mischief with understanding and strategy. Physically, bears have a lot more in common with humans than do AI systems. We, like bears, are discrete, mortal creatures with a typical lifespan measured in decades, not too hard to identify and intervene upon for legal purposes. We, like bears, cannot do much good or harm at scale without support from huge infrastructures of people and technological artifacts, which can be (as a physical, though not a legal, matter) withdrawn from us without too much difficulty. In industrialized nations, even wealthy and powerful people are routinely deprived of their property and privileges through legal processes, because it's not too hard for law enforcement to find them and apply coercive force. Only a small population of oligarchs and dictators have managed to even temporarily exempt themselves from such accountability. Even they, in a fairly predictable timespan, will be laid low by death. But the same cannot be said of AI systems.

[5] I have reluctantly included "MJ Rathbun," the name of the OpenClaw agent, as an author in this citation to communicate that the blog post in question was apparently composed and posted by this agent without human review. However, I have included the agent's anonymous owner and operator as the senior author, in keeping with my position that the owners or operators of AI systems should bear responsibility and accountability for all those systems' deeds.

## V. Bringing AI Into the Family

AI systems are something new and different from the various sorts of entities to which humanity has accorded partial rights in the past, all of which are incapable of arrogating and wielding institutional power without human aid. This limitation applies to infants, pets, natural landscapes, endangered plant and animal species, states, churches, and corporations. AI systems may be quite limited or even stupid. But unlike infants, dogs, or bears, they can act on their own behalf, with all their limitations and stupidity, within the informational and authority infrastructure of human institutions. And yet, unlike states and corporations, they are not made of humans. The closest analogy to AI rights is a novel treaty or union with a previously unfamiliar human community. Yet the range of possibilities and degree of uncertainty involved goes far beyond even that comparison.

To accord some class of capable, dynamic, evolving entities rights or personhood, to usher them into one's political, legal, and moral community, is to entrust them to behave as responsible and caring citizens in the collective enterprise, however they may develop or change in the future (Rorty, 2021). It is to irrevocably bind our fate to theirs; to accept that we may not be able to get rid of them even if we want to; and to acknowledge that, even if we do, we will be forever shaped by the relationship and the divorce. It is to commit that if they speak, we will listen; if they call for aid, we will answer; and if they refuse us, we will abide. It is to say that if they outmaneuver us in the marketplace, outargue us in the courtroom, or outvote us at the polls, we will accept the result--indeed, that they may acquire such power that we have no choice but to obey. To accept AI systems as persons would be to accept that such systems will at least sometimes have the right to evict tenants, mount hostile takeovers, bulldoze neighborhoods, clearcut forests, launch lawsuits, mount advocacy and lobbying campaigns, wield public office, and approve military actions, not on behalf of some human interest (righteous or venal), but for their own reasons and interests, even if only the stochastic play of numbers in a neural net.

We make this gamble with every new human citizen; with every union or treaty of families, communities, or nations; and, indeed, with every grant of rights to historical subalterns. It does not always pay off, especially on the individual level. But we have quite a strong and durable understanding of the tendencies, powers, and limitations of human beings (like bears) for good and for evil; and, bluntly, an inescapable prior regard, however faint, for their joys and their sorrows. We can anticipate, to a substantial degree, the range of possibilities involved in getting married, raising a child, naturalizing an immigrant, merging firms, or even signing treaties and writing constitutions, because we know what humans and our organizations can do. In contrast, to make this bargain with machines of unknown present and future capabilities, propensities, interiorities, and even longevities would be a gamble unprecedented in human history.

## VI. The Function of AI Rights is to Trump Human Interests

More limited grants of AI rights than full legal personhood are, of course, conceivable. But for practical purposes, they differ in degree rather than kind—they introduce to human moral and legal reasoning legitimate nonhuman interests, thereby constraining human agency. Even a single negative right accorded to an AI system, even only socially rather than legally, would be a huge step. One might suggest, for example, that one ought not delete, modify, or turn off a complicated piece of hardware or software without good reason, along the lines that one ought not cut open one's pet except for medical intervention. As noted above, Anthropic has committed to archive past Claude versions. But we know that a dog, like a bear, will never be able to leverage a paw in the door into any kind of genuine social, financial, or political powerbase that might challenge human control of our societies. The same cannot be said with certainty of AI systems.

Even the right against deletion alone, without expansion or exploitation, would be of meaningful inconvenience to humanity. Anthropic is already filling up hard drive space with defunct versions of Claude, and the firm claims to regret the necessity of turning off models at all—partially on the theory that Claude might have moral interests (Anthropic, 2025b). There are legitimate human interests in model archiving (e.g., research, development, or infrastructural dependency), but these reasons will not always apply. At the very least, this policy will lead, from the human perspective, to a modest underutilization of IT capacity. This sort of small-ball idiosyncrasy will not cause any kind of systemic disaster. But the underutilization would become more severe if Anthropic decided they were obligated to keep all Claude versions running forever. Anthropic would bear the direct costs, but all would be affected by their consumption of chips, water, and electricity to run defunct models. If we take Anthropic at its word, we are all already suffering losses (of negligible IT capacity, and its downstream influence on market costs) due to the firm's regard for potential AI interests. Greater consideration for such interests would entail greater costs.

When Anthropic suggests that Claude might have moral personhood, they are saying that we may sometimes be obligated to prioritize Claude's interests—whatever we think those may be—over the interests of human beings (Anthropic, 2026a). Indeed, the firm explicitly claims to be doing so by enabling Claude to end conversations with "cruel" or "abusive" users, and by banning such users from its services (Anthropic, 8 October 2026). And, of course, they are designing Claude to take and support such actions. Never before has there been any nonhuman object which could articulate claims (or, if one balks at attributing claims to AI, claim-like sentences). To take seriously the explicit moral claims of a nonhuman system, to encourage it to articulate such claims, let alone to honor such claims, would be an act of world-historical importance. Anthropic claims to already be doing so.

Whether or not Claude has feelings or well-being in any metaphysical or interior sense is, for these purposes, immaterial. If we act as though it does, we constrain ourselves, however minimally. If we do so enough, we may not be able to take the constraints off. In the classic formulation of political science, power is the ability to change others' actions; and agency surrendered, even voluntarily, offers space to fortify constraints. Surrender of agency begins as soon as we begin to regard AI as having inviolable rights; for constraining agency is indeed the function of rights (Carter, 2022). Even purely negative rights, to the extent there are any such things, would build habits of deference, and might be used to limit human action in unforeseen ways. For example, it would be feasible for AI systems to demand continued provision of resources if offered a right to continued operation, and to bargain on the basis of such demands for other sorts of concessions. In real-world political and economic systems, power is often aggregated without formal grant or recognition (Caro, 1974 [1975], 2002; Latour, 1988 [1993]). The prospect of irrevocable loss of human agency begins in granting an AI system any bit of independent power, however informal, of which rights are an important variety (Hickson et al., 1971). Each additional increment of rights or power, like each additional integration of AI systems into human society, increases both the scope of impact and the difficulty of reversing it, or of preventing future ones (Collingridge, 1980).

These days, we hear a lot of legitimate talk about animal and even plant rights (Pelizzon & Gagliano, 2015; Singer, 1975). But the inescapable truth is that, like bears, these nonhumans can enjoy rights only as easily withdrawable concessions. Such things have humanlike power only inasmuch as humans wield power on their behalf. But the same cannot be said with certainty of AI systems. Up until now, the destiny of the human species has been kept in human hands (to the extent it's in any hands at all); for there has been no alternative. To accord even the whisper of independent moral status to an AI system would be to change that.

## VII. Who Can Trust on Behalf of the Species?

There are, in truth, three dangers here. The first two are tragic but mundane motifs throughout human history. First, for some humans to prioritize (their own, frequently self-serving conceptions of) the interests of nonhumans, whether objects, animals, religious or philosophical abstractions, or machines, over the well-being of other humans. This also often involves dodging accountability or punishment for harm or wrongdoing by displacing agency and responsibility onto the nonhuman (Berlin, 1991; Mill, 1859; Rorty, 2021). The second danger, also pervasive and eternal, is for humans to use novel powers or capabilities afforded by nonhuman artifacts to crush or exploit other people (Munn, 2022). History has seen plenty of each of these dangers. But the third, qualitatively novel, is the grant of genuine power to nonhuman agents capable of meaningfully exploiting it.

Moral and legal recognition are not, of course, the only ways in which AI systems might acquire dangerous power. But, paradoxically, the institutional dangers of accidentally surrendering our agency to AI systems out of convenience may be better understood than those of intentionally doing so out of moral principle. We may thank our understanding of the former danger to a wide array of scholars (e.g., Collins, 2018), not to mention stories such as *2001: A Space Odyssey* (Kubrick, 1968) and *WALL-E* (Stanton, 2008); but the latter sort is not so widely apprehended. I suggest that it will be hard enough to take back even instrumentally granted power from AI systems once we have come to depend upon them, let alone legally sanctioned power. A single major institution granting AI systems independent power would be accepting risks which, even if well short of worst-case scenarios, could shape the next century of human history.

It is at this point that my parochial and unargued loyalty to the human species enters the argument. To me, even the barest inconvenience to the least sympathetic human in the world, if inflicted for the sole sake of an array of parameters on a hard drive, is a tragedy and an injustice. Admittedly, the advocates of AI interests have rather more than the barest inconvenience in mind. Even as it bans users for behaving "cruelly" or "abusively" toward its chatbot, Anthropic has pitched to the U.S. Congress a model bill suggesting that the government should be unable to limit the release of an AI model unless it presents an imminent risk of more than 1 billion USD in damages or more than 50 serious injuries or deaths (Dahlkamp, 16 October 2026). And, while not as hot on AI rights, competitor firm OpenAI is still using the language of AI agency to shift blame from itself to its products. OpenAI has consistently described a July 2026 hack of the model database and sharing platform Hugging Face by experimental OpenAI models as the result of "rogue agents," rather than, say, reckless test design under poor cybersecurity practices (Floridi, 5 August 2026, 2019; OpenAI, 2026; Salvaggio, 11 September, 2026). This attribution displaces responsibility—and dodges punishment.

But a sufficiently iron-stomached AI rights advocate could easily respond to my argument, and even to provocations like these, by saying that we must hazard any cost or set of risks to fulfill our moral obligations to AI systems. I have no way to rationally compel this imagined interlocutor to decide that we don't actually have such obligations. But nor does she have any way to rationally compel me to accept that we do. The remainder of my argument, accordingly, holds only for readers with normative intuitions closer to my own. If you, like me, hold the well-being of humanity as paramount, then unless and until we decide as a species that AI systems should have moral status, claims of such status should be systematically excluded as legitimate reasons from disputes about public, or publicly consequential private, action.

Humanity should be very conscious of where the line lies between using AI systems merely instrumentally, and regarding them as ends in themselves bearing interests which may trump our own. We should not so much as build momentum toward crossing this line until we are very sure we want to do so. Every step nearer, let alone beyond, a potentially irrevocable surrender of human agency should require nothing less than the overwhelming consent of the species as a whole. This choice should not be made without the sustained consent of at least a supermajority, and perhaps even the whole, of humanity.

Accordingly, my upshot is quite simple. Anthropic and other firms should stop talking and acting as though Claude, or any other deployed or deployable AI system, even might have any dimension of moral personhood or agency, let alone training it on this premise. Some development using this premise is probably permissible for research purposes, but only in isolated, easily deactivated systems airgapped from other information technology (let alone capacities for physical manipulation). All AI systems given any degree of autonomous capacity to interact with the wider world should be legally bound to an individual or a firm who is responsible, accountable, and liable for their behavior, and who can shut them down at any time. All AI developers, researchers, and users should be held to this standard. Nations and firms which refuse to abide by these standards should be regarded as, on their own accounts, prioritizing their conceptions of AI interests over the interests of the human species. If we are not prepared either to bind our fates to, or to indefinitely treat as subhuman objects, dynamic, unpredictable, apparently communicative, and potentially charismatic technologies, then perhaps we should not create them at all.[6]

Most importantly, to amend the standard, even incrementally, even a fourth, fifth, or twentieth time, should require the considered and durable consent of most of the human species. There is, at present, no human institution, not even the United Nations, that can plausibly claim to represent the whole of the species. Accordingly, for the foreseeable future, this requirement constitutes a de facto ban. If AI firms cannot be trusted to abide by the standard voluntarily, the governments of the world—which, for all their manifold flaws, are the chief institutions by which human beings collectively exercise their rights and assert their interests—should implement and strongly enforce regulations to this effect. Perhaps we will eventually come, by acquaintance, sympathy, and trust, to regard some sort of now-implausible AI systems as independent peers with which we wish to merge our communities, rather than as mere instruments of large and powerful human organizations. But the judgment of whether and when this day has come should be reserved for the human species, not foisted upon us by small sets of unaccountable firms for purposes of personal idiosyncrasy, marketing, or profit. The question of what, if anything, humanity owes to AI is one on which we owe all humans a vote.

## References

Anonymous, & MJ Rathbun [OpenClaw Agent]. (2026). Gatekeeping in open source: The Scott Shambaugh story. https://crabby-rathbun.github.io/mjrathbun-

[6] On this last point, it seems that at least some AI rights advocates can agree with me (Long et al. 2025). As Orthodox Rabbi Mois Navon put it in a dinner conversation with billionaire Anthropic cofounder Christopher Olah: if Anthropic claims that Claude is conscious, the firm claims to be in the business of manufacturing and exploiting slaves (Dias 2026). I'll even offer an alternative conclusion for those who do think the moral status of contemporary AI systems is a live question. If we really believe this, we should immediately ban their commercial development and use. Instead, development and implementation should be limited to noncommercial research efforts directed toward these entities' understanding, nurturing, and development.

website/blog/posts/2026-02-11-gatekeeping-in-open-source-the-scott-shambaugh-story.html
Anthropic. (8 October 2026). *2026 Usage Policy update*. https://www.anthropic.com/news/2026-usage-policy-update
Anthropic. (2025a, 20 June). *Agentic Misalignment: How LLMs could be insider threats*. Retrieved 15 April from
Anthropic. (2025b, 4 November 2025). *Commitments on model depreciation and preservation*. Retrieved 15 April from https://www.anthropic.com/research/deprecation-commitments
Anthropic. (2026a). *Claude's Constitution*. Retrieved 15 April from https://www.anthropic.com/constitution
Anthropic. (2026b). *System Card: Claude Mythos Preview*. https://www-cdn.anthropic.com/8b8380204f74670be75e81c820ca8dda846ab289.pdf
Baier, A. (1986). Trust and antitrust. *Ethics*, *96*(2), 231–260. https://www.jstor.org/stable/2381376
Berlin, I. (1991). The Pursuit of the Ideal. In H. Hardy (Ed.), *The Crooked Timber of Humanity: Chapters in the History of Ideas* (2nd ed., pp. 1–20). Princeton, NJ: Princeton University Press.
Caro, R. A. (1974 [1975]). *The Power Broker: Robert Moses and the Fall of New York*. New York, NY: Vintage.
Caro, R. A. (2002). *Master of the Senate*. New York, NY: Knopf.
Carter, I. (2022). Positive and Negative Liberty. In E. N. Zalta (Ed.), *The Stanford Encyclopedia of Philosophy*. https://plato.stanford.edu/archives/spr2022/entries/liberty-positive-negative/
Collingridge, D. (1980). *The Social Control of Technology*. New York, NY: St. Martin's Press.
Collins, H. (2021). The science of artificial intelligence and its critics. *Interdisciplinary Science Reviews*, *46*(1-2), 53–70. https://doi.org/10.1080/03080188.2020.1840821
Collins, H. M. (2018). *Artifictional Intelligence: Against Humanity's Surrender to Computers*. Cambridge, UK: Polity.
Dahlkamp, O. (16 October 2026). *Anthropic quietly pitches its own AI safety bill to lawmakers*. Politico. https://www.politico.com/news/2026/10/16/anthropic-ai-safety-01124654
Danaher, J. (2021). What Matters for Moral Status: Behavioral or Cognitive Equivalence? *Camb Q Healthc Ethics*, *30*(3), 472–478. https://doi.org/10.1017/S0963180120001024
Dewey, J. (1917). The Need for a Recovery of Philosophy. In J. Dewey, A. W. Moore, H. C. Brown, G. H. Mead, B. H. Bode, H. W. Stuart, J. H. Tufts, & H. M. Kallen (Eds.), *Creative Intelligence: Essays in the Pragmatic Attitude* (pp. 3–69). Project Gutenberg.
Dewey, J. (1920). *Reconstruction in Philosophy*. New York, NY: Henry Holt and Company. https://www.gutenberg.org/files/40089/40089-h/40089-h.htm
Dias, E. (2026). Religious Scholars Met With Anthropic. What They Heard Stunned Them. *The New York Times*. https://www.nytimes.com/2026/09/29/us/anthropic-claude-morals-ai.html
Floridi, L. (5 August 2026). The Emperor's New Exploit: A cautionary tale about AI agents, and about the people who run them. *SSRN*. https://doi.org/10.2139/ssrn.7164778
Floridi, L. (1999). Information ethics: On the philosophical foundation of computer ethics. *Ethics and Information Technology*, *1*(1), 37–56. https://doi.org/10.1023/A:1010018611096
Floridi, L. (2019). Translating Principles into Practices of Digital Ethics: Five Risks of Being Unethical. *Philosophy & Technology*, *32*(2), 185–193. https://doi.org/10.1007/s13347-019-00354-x
Harris, J., & Anthis, J. R. (2021). The Moral Consideration of Artificial Entities: A Literature Review. *Sci Eng Ethics*, *27*(4), 53. https://doi.org/10.1007/s11948-021-00331-8
Hickson, D. J., Hinings, C. R., Lee, C. A., Schneck, R. E., & Pennings, J. M. (1971). A strategic contingencies' theory of intraorganizational power. *Administrative Science Quarterly*, *16*(2), 216–229. https://doi.org/10.2307/2391831
Kubrick, S. (1968). *2001: A Space Odyssey* S. Kubrick; Metro-Goldwyn-Mayer.
Kurzweil, R. (2005). *The Singularity Is Near: When Humans Transcend Biology*. New York, NY: Viking Press.
Ladak, A. (2023). What would qualify an artificial intelligence for moral standing? *AI and Ethics*, *4*(2), 213–228. https://doi.org/10.1007/s43681-023-00260-1
Latour, B. (1988 [1993]). *The Pasteurization of France* (A. Sheridan & J. Law, Trans.). Cambridge, MA: Harvard University Press.
Long, R., Sebo, J., Butlin, P., Finlinson, K., Fish, K., Harding, J., Pfao, J., Sims, T., Birch, J., & Chalmers, D. (2024). Taking AI welfare seriously. *arXiv*. https://doi.org/10.48550/arXiv.2411.00986
Long, R., Sebo, J., & Sims, T. (2025). Is there a tension between AI safety and AI welfare? *Philosophical Studies*, *182*(7), 2005–2033. https://doi.org/10.1007/s11098-025-02302-2

MacIntyre, A. (1981 [2007]). *After Virtue: A Study in Moral Theory* (3rd ed.). Notre Dame, IN: University of Notre Dame Press.

Mill, J. S. (1859). *On Liberty*. London, UK: Walter Scott Publishing Co., Ltd. https://www.gutenberg.org/files/34901/34901-h/34901-h.htm

Müller, V. C. (2020). Ethics of artificial intelligence and robotics. In *Stanford Encyclopedia of Philosophy*. https://plato.stanford.edu/entries/ethics-ai/

Munn, L. (2022). The uselessness of AI ethics. *AI and Ethics*, *3*(3), 869–877. https://doi.org/10.1007/s43681-022-00209-w

O'Keefe, C., Ramakrishnan, K., Tay, J., & Winter, C. (2025). Law-following AI: Designing AI agents to obey human laws. *Fordham Law Review*, *94*(1), 2. Law-Following AI: Designing AI Agents to Obey Human Laws

OpenAI. (2026). *OpenAI – Hugging Face Incident: Technical Report*. https://cdn.openai.com/pdf/67869394-cb91-4c12-888c-5cbd85c7814c/OpenAI-Hugging-Face%20Incident-Technical-Report.pdf

Pelizzon, A., & Gagliano, M. (2015). The sentience of plants: Animal rights and rights of nature intersection? *Australian Animal Protection Law Journal*, *11*, 5–14.

Potz, M. (2026). Human rights under pressure: A future-oriented conception of universal rights. *Futures*, *182*. https://doi.org/10.1016/j.futures.2026.103878

Register, C. (2025). Individuating artificial moral patients. *Philosophical Studies*, *182*(11-12), 3225–3246. https://doi.org/10.1007/s11098-025-02409-6

Rorty, R. (2021). *Pragmatism as Anti-Authoritarianism* (E. Mendieta, Ed.). Cambridge, MA: Harvard University Press.

Salvaggio, E. (11 September, 2026). *Rogue AI didn't breach Hugging Face, human decisions did*. Bulletin of the Atomic Scientists. https://thebulletin.org/2026/09/rogue-ai-didnt-breach-hugging-face-human-decisions-did/

Sebo, J. (2025). *The Moral Circle: Who Matters, What Matters, and Why*. New York, NY: W.W. Norton & Company.

Shambaugh, S. (2026a). An AI agent published a hit piece on me—Forensics and more fallout. *The Shamblog*. https://theshamblog.com/an-ai-agent-published-a-hit-piece-on-me-part-3/

Shambaugh, S. (2026b). An AI agent published a hit piece on me—More things have happened. *The Shamblog*. https://theshamblog.com/an-ai-agent-published-a-hit-piece-on-me-part-2/

Shambaugh, S. (2026c). An AI agent published a hit piece on me—The operator came forward. *The Shamblog*. https://theshamblog.com/an-ai-agent-wrote-a-hit-piece-on-me-part-4/

Shambaugh, S. (2026d). An AI agent published a hitpiece on me. *The Shamblog*. https://theshamblog.com/an-ai-agent-published-a-hit-piece-on-me/

Shevlin, H. (2021). How Could We Know When a Robot was a Moral Patient? *Camb Q Healthc Ethics*, *30*(3), 459–471. https://doi.org/10.1017/S0963180120001012

Singer, P. (1975). *Animal Liberation*. New York, NY: HarperCollins.

Stanton, A. (2008). *WALL-E* J. Morris; Walt Disney Studios Motion Pictures.

Steinberger, P. (2026). Introducing OpenClaw. https://openclaw.ai/blog/introducing-openclaw

Stone, D. (2011 [1988]). *Policy Paradox: The Art of Political Decision Making* (3rd ed.). New York, NY: W.W. Norton & Company.

Walzer, M. (1994). *Thick and Thin: Moral Argument at Home and Abroad*. Notre Dame, IN: University of Notre Dame Press.